%% file: martayan.tex
\documentclass[multphys,vecphys]{svmult}

\usepackage{graphicx}        
\usepackage{multicol}        
\usepackage[bottom]{footmisc}

\begin{document}

\title*{ALBUM: a tool for the analysis of slitless spectra and its application
to ESO WFI data.}
\titlerunning{ALBUM: a tool to WFI slitless spectra} 
\author{C. Martayan\inst{1,2} 
\and D. Baade\inst{1} 
\and A.-M. Hubert\inst{2}
\and M. Floquet\inst{2} 
\and J. Fabregat\inst{3} 
\and E. Bertin\inst{4}}
\authorrunning{C. Martayan, D. Baade et al.} 
\institute{European Organisation for Astronomical Research in the Southern Hemisphere, Karl-Schwarzschild-Str. 2, 85748 Garching b. Muenchen, Germany
\texttt{dbaade@eso.org}
\and GEPI-Observatoire de Paris, 5 place Jules Janssen, 92195 Meudon cedex, France \texttt{christophe.martayan@obspm.fr}
\and Observatori Astron\`omic de la Universitat de Val\`encia, Edifici Instituts
d'Investigaci\'o, Pol\'igon La Coma, 46980 Paterna Val\`encia, Spain
\and Institut d'Astrophysique de Paris, 98bis boulevard Arago, 75014 Paris, France
}
%
%
\maketitle

\begin{abstract}
ALBUM is a general-purpose tool to visualize and screen large amounts of slitless spectra.   It was developed
for a search for emission-line stars in SMC and LMC clusters.   The observations were obtained with ESO's
Wide Field Imager (WFI) and comprise $\sim$8 million low-resolution spectra.   The tool as well as the results of
its application to the SMC part of the database are presented.   The inferred frequency of Be stars is
compared to the one in the higher-metallicity environment of the Milky Way. 
\end{abstract}

\begin{figure}
\centering
\includegraphics[height=12cm, angle=-90]{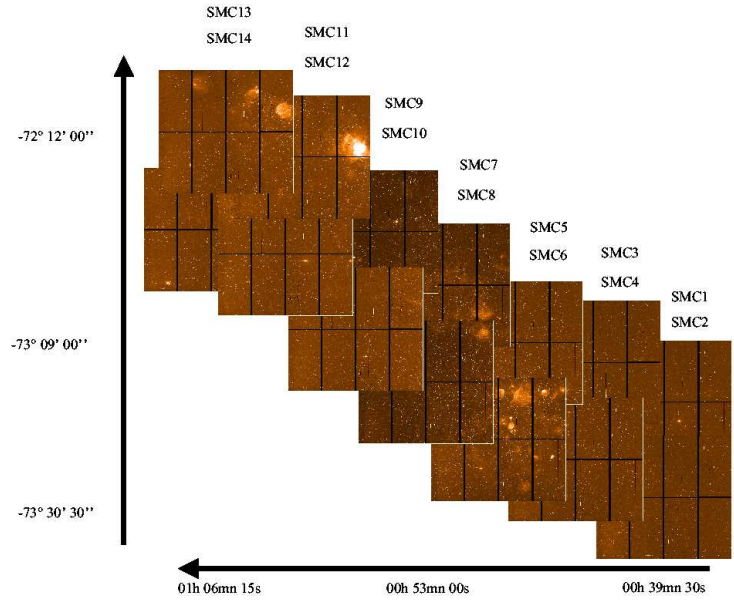}
\caption{The Small Magellanic Cloud as projected on the WFI frames used in this study.}
\label{fig1}       
\end{figure}

\begin{figure}
\centering
\includegraphics[height=3cm]{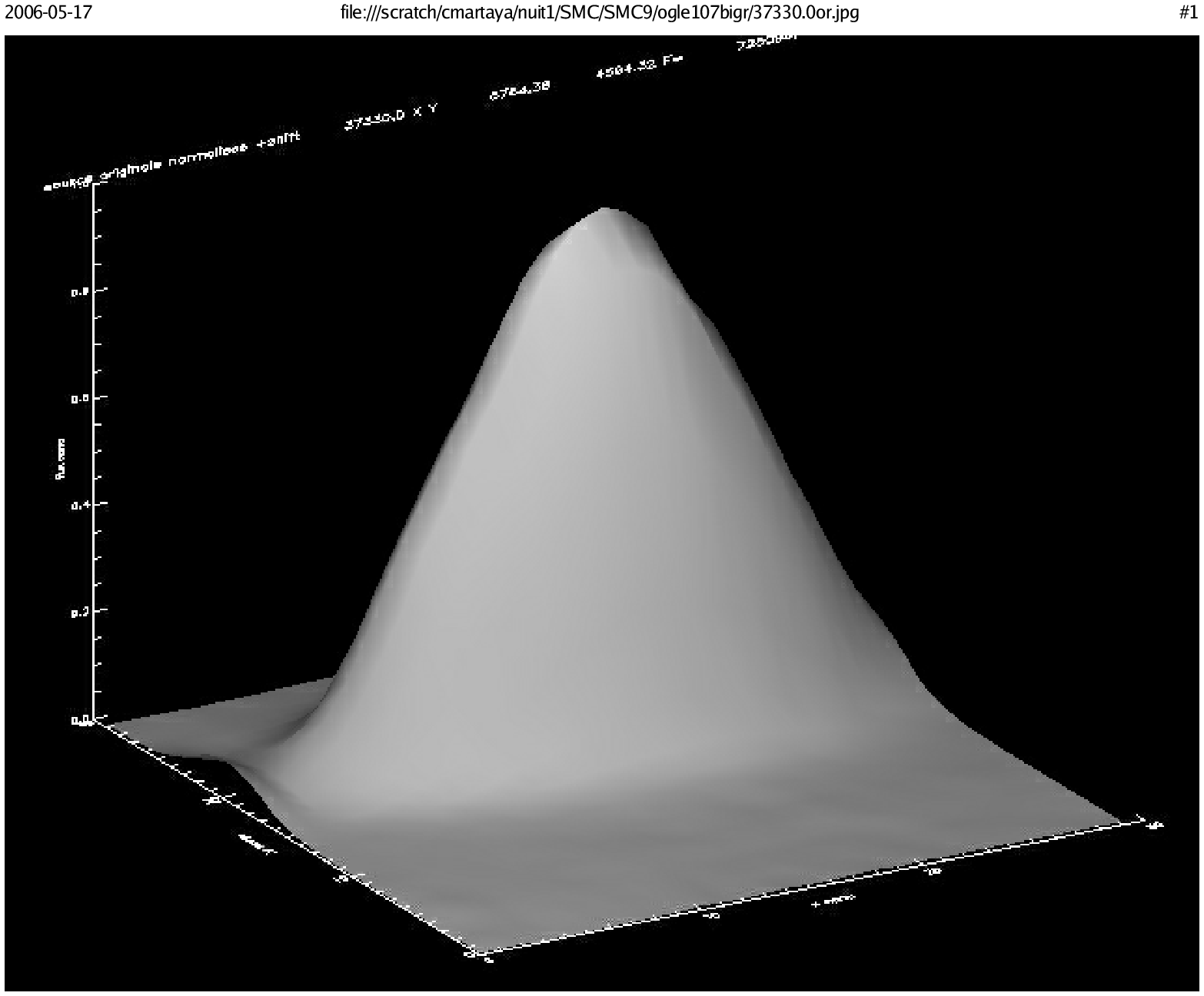}
\includegraphics[height=3cm]{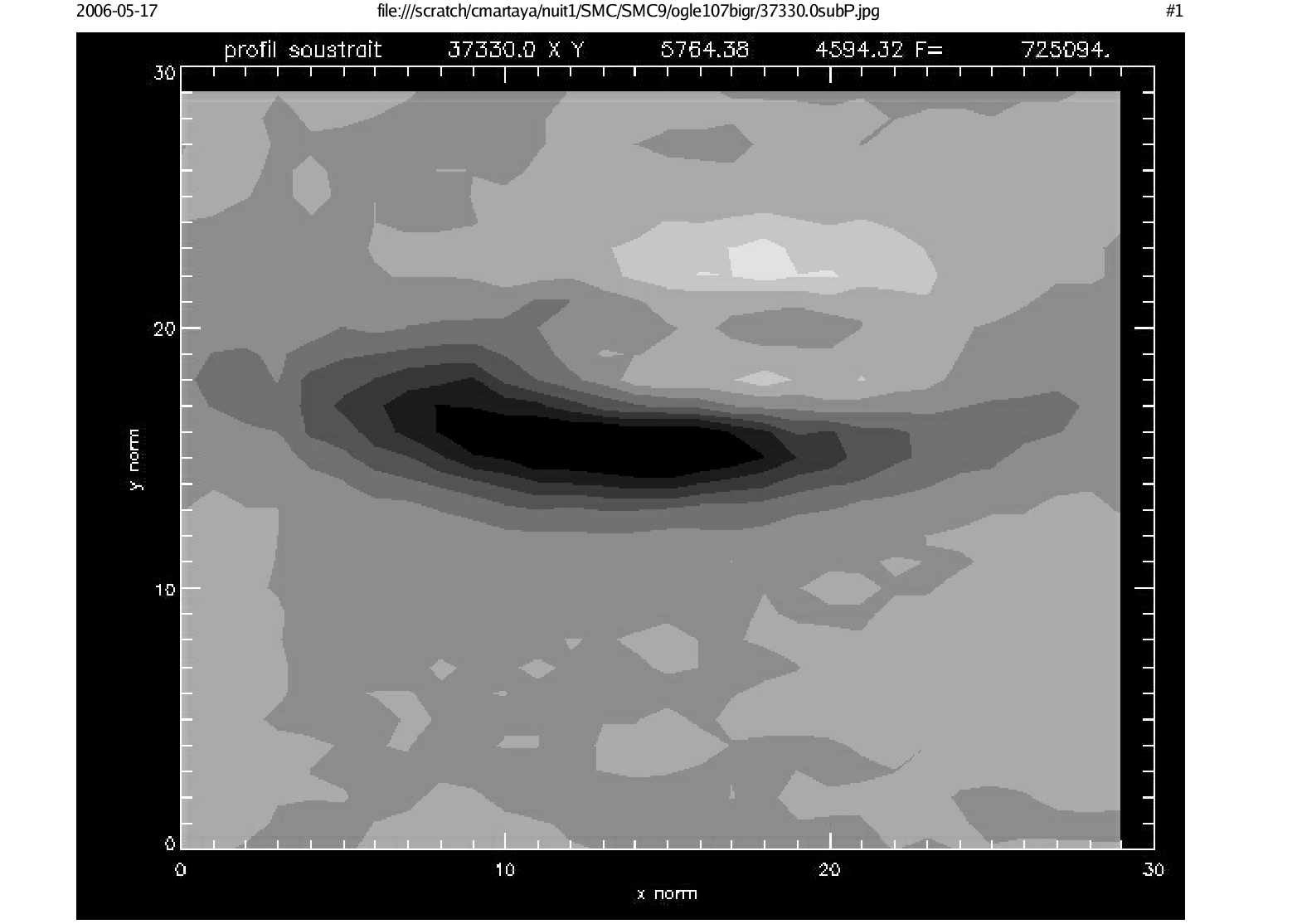}
\caption{Non Emission line star. Left panel: original source. Right panel: projection of the residual 
of the subtraction of the mean profile}
\label{fig2}       
\end{figure}

\begin{figure}
\centering
\includegraphics[height=3cm]{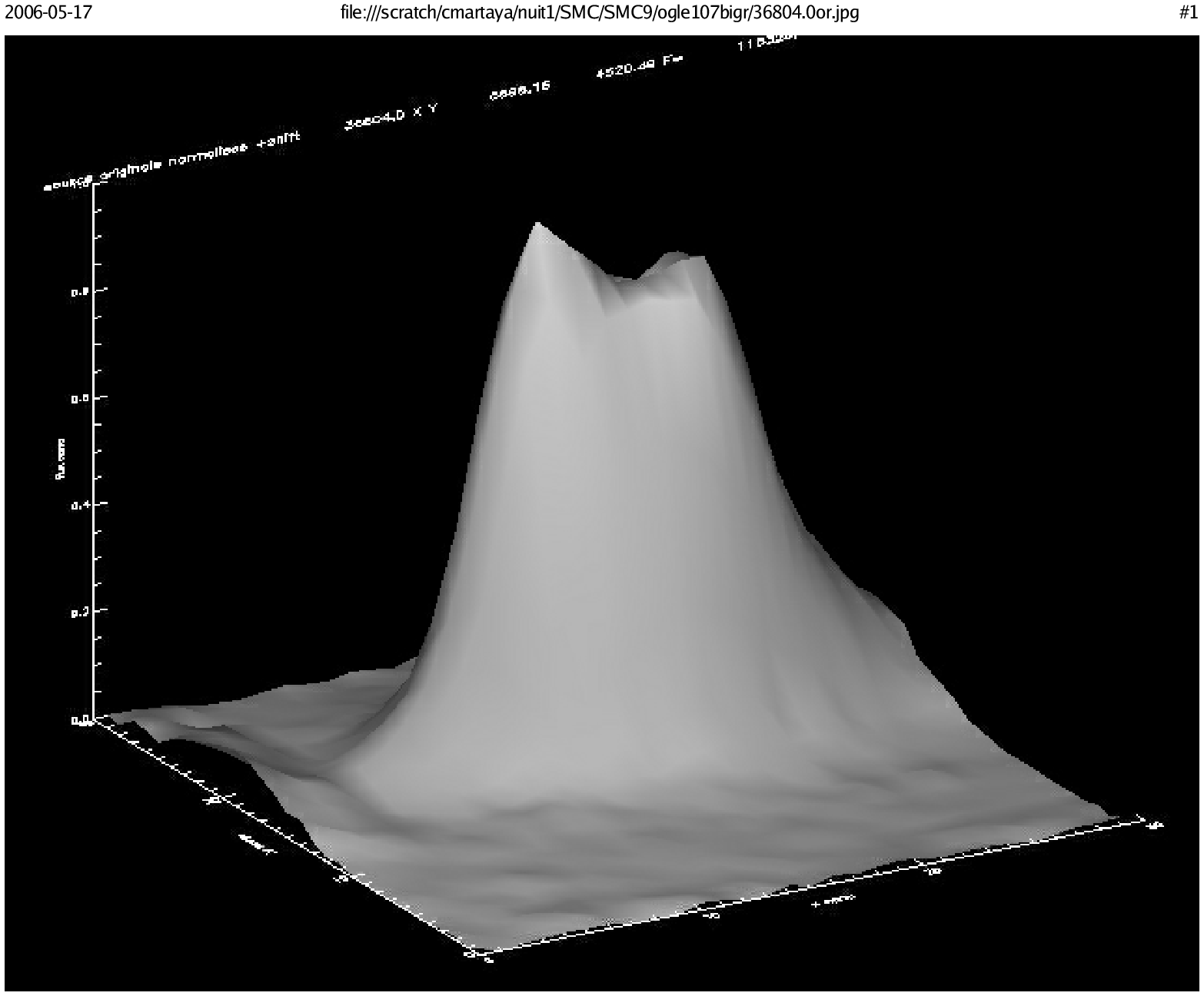}
\includegraphics[height=3cm]{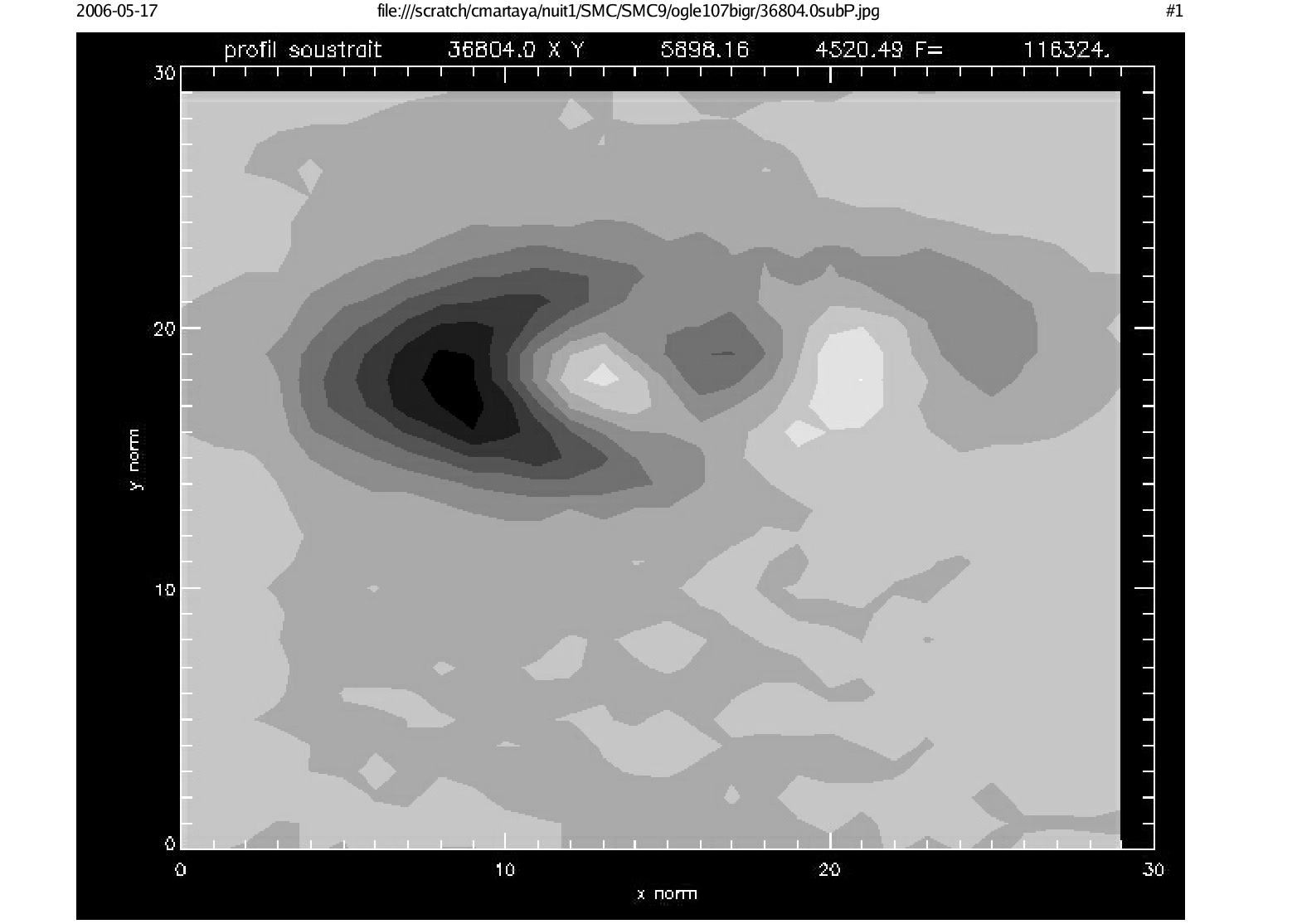}
\caption{Emission line star. Left panel: original source. Right panel: projection of the residual 
of the subtraction of the mean profile, due to the emission in H$\alpha$ and the defocus 
a ring structure with 2 peaks is observed.}
\label{fig3}       
\end{figure}

\section{Data reduction}

Observations (see Fig.~\ref{fig1}) covering much of the Small Magellanic Cloud (SMC) have been obtained in September 2002
with the WFI attached to the 2.2-m MPG Telescope at La Silla.  The instrument was used in its slitless
spectroscopic mode. To reduce crowding, the length of the spectra was limited by means of a filter with a
bandpass of 7.4 nm centered on H$\alpha$. Unfortunately, a large part of the fields suffers from substantial
non-homogeneous defocusing,  which severely reduces the contrast between stars with and without line emission
at H$\alpha$.

The basic reduction of the CCD images was performed with the  MSCRED IRAF tasks except for the astrometry,
for which the ASTROM package (\cite{WG03}) was applied to the extracted 1st-order spectra. The achieved
accuracy was 0.5-1'' rms.  The extraction in 2-D of the spectra was accomplished by means of the
SExtractor software (\cite{bertin96}).  All in all, about 1 million of the 3 million spectra available in the
SMC part of the  survey proved usable.

To recognize and distinguish emission-line stars (Em**) from other objects,  we created the ALBUM package in
idl.   Its strategy is based on the assumption that the 2-D point-spread function (PSF) is only slowly
varying with position in the frame and only insignificantly falsified  by the inclusion of emission-line
objects in the calculation of the mean local PSF.   Typically, 50-250 spectra were aligned (by cross
correlation), co-added, and normalized.   This local template spectrum was subtracted (after cross
correlation and shift) from each  normalized 2-D spectrum (see Fig.~\ref{fig2}) to be checked for H$\alpha$ line emission.  In the
case of Em*,  the 2-D spectra show a secondary peak (see Fig.~\ref{fig3}).  But after subtraction of the mean PSF the resulting 
difference images display a more characteristic and conspicuous ring-like structure,  which is due to the
large defocus (see Fig.~\ref{fig3}).  Since this peculiar structure is more readily and  reliably recognized by the human eye than
by software, the identification of the Em* was  done by visual inspection of the album of PSF-subtracted 2-D
spectra.

\section{Results: frequency of Be stars vs. metallicity}

\begin{figure}
\centering
\includegraphics[height=12cm, angle=-90]{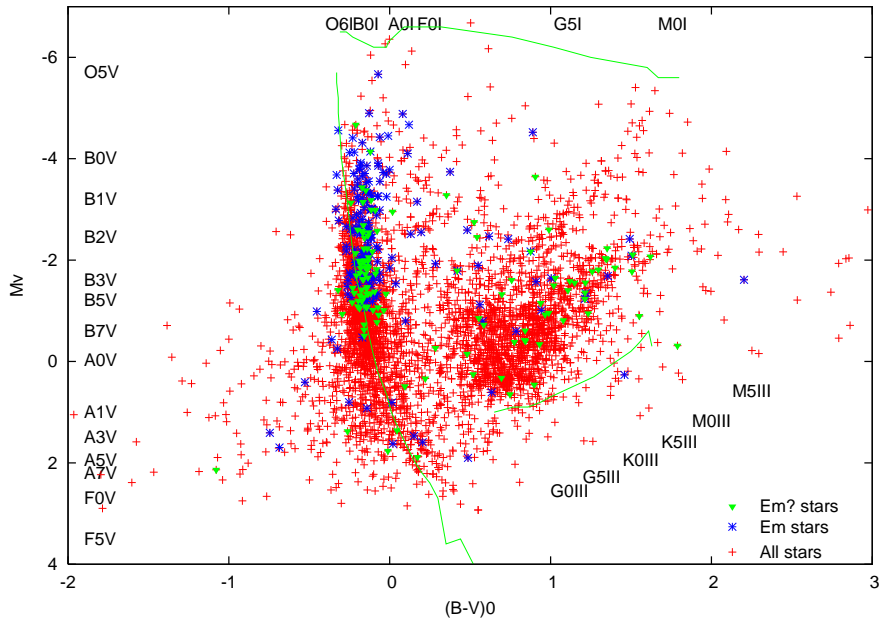}
\caption{(Mv,(B-V)$_{O}$ diagram for the stars cross-correlated in OGLE database. The calibration in spectral
types comes from \cite{Lang91}. Red '+' are for the non-Emission line stars, Blue '*' for the Emission line
stars, and the green triangles are for the candidate Emission line stars.}
\label{fig4}       
\end{figure}

We have investigated 85 clusters in the SMC with log(age) between 7 and 9 and E[B-V] available from the OGLE survey
(\cite{pi99}). For a total of 7741 stars, V, B, and I magnitudes were obtained from the OGLE database (\cite{pi99},
\cite{szy05}). Fig.~\ref{fig4} displays the combined HR diagram of all clusters with the Em** marked. The results can
be compared to the relative  frequencies of Be stars (Be/(B+Be)) in Milky Way clusters (\cite{MCS05}) in order to 
search for any effect of the metallicity on the proportion of Be stars and on the still unknown  reason for the
development of disks around these extremely rapidly rotating stars.   The fractions range from 0 to 46\% in the SMC
and from 0 to 24\% in the Milky Way (MW), depending on the parent cluster.   There seems to be a trend in that the
lower the metallicity, the higher the proportion of Be stars is.   This could be explained by higher rotational
velocities in the SMC than in the MW  (\cite{MM01}, \cite{marta07}). 

\section{Conclusions}
\begin{itemize}
\item A new method and software package for the reduction and analysis of slitless spectra was developed. 
\item It was applied to WFI data.  Catalogues of Be stars in 85 SMC clusters were obtained.
\item Metallicity seems to influence the relative proportion of Be stars among all B-type stars:  
The fraction of Be stars is larger in the SMC than in the MW.
\item The equivalent study of the LMC is in progress (additional 5 million spectra).
\item The combination of a large field and slitless spectroscopy is very powerful for surveys for objects 
with distinct spectral properties.
\end{itemize}

\begin{acknowledgement}
C.M. gratefully acknowledges support from ESO's DGDF 2006.
\end{acknowledgement}

%
%
%
\input{refmartayan}



\end{document}

%% file: refmartayan.tex
%
%

%
%

%% file: martayan.bbl
\begin{thebibliography}{99.}
%
%
%









\bibitem{bertin96} E. Bertin, S. Arnouts: A\&AS \textbf{117}, 393 (1996)

\bibitem{Lang91} K. R. Lang: \textit{Astrophysical data: Planets and stars}, Eds
Springer-Verlag, New York (1992)

\bibitem{MM01} A. Maeder, G. Meynet: A\&A \textbf{373}, 555 (2001)

\bibitem{marta07} C. Martayan, Y. Fr\'emat, A.-M. Hubert, et al.: A\&A \textbf{462}, 683 (2007) 

\bibitem{MCS05} M. V. McSwain, D. Gies: ApJS \textbf{161}, 118 (2005)

\bibitem{pi99} G. Pietrzy\'nski, A. Udalski, M. Kubiak, et al.: AcA \textbf{49}, 521 (1999)

\bibitem{szy05} M. Szymanski: AcA \textbf{55}, 43 (2005)

\bibitem{WG03} P. T. Wallace, N. Gray: \textit{User's guide of ASTROM} (2003)






\end{thebibliography}
